\documentstyle[aps,multicol]{revtex}
\begin{document}
\newtheorem{lemma}{Lemma}
\newtheorem{thm}{Theorem}
\newtheorem{cor}{Corollary}
\newtheorem{defn}{Definition}
\rightline{(physics/9801010)}
\centerline{\Large{\bf Local fractional derivatives and}}
\centerline{\Large{\bf fractal functions of several variables}} 

\vspace{10pt}

\centerline{ Kiran M. Kolwankar\footnote{email: kirkol@physics.unipune.ernet.in} 
and Anil D. Gangal\footnote{email: adg@physics.unipune.ernet.in}}
\vspace{3pt}

\centerline
{\it Department of Physics, University of Pune, Pune 411 007, India.}
\vspace{6pt}
\begin{abstract}
The  notion of a local fractional derivative (LFD) was introduced recently
for functions of a single variable.
 LFD was shown to be useful
in studying fractional differentiability properties of fractal
and multifractal functions.
It was demonstrated that the local H\"older exponent/
dimension was directly related to the maximum order for which LFD existed.
We have extended this definition to directional-LFD
 for functions of many variables and
demonstrated its utility with the help of simple examples.
\end{abstract}

\begin{multicols}{2}
\section{Introduction} \label{secintro}
Fractal and multifractal functions and 
the corresponding curves or surfaces are found in numerous 
places in nonlinear and nonequillibrium phenomenon.
For example, isoscalar surfaces for advected scalars in certain turbulence problems
\cite{19,20},
typical Feynman \cite{30,31} and Brownian paths \cite{2,10},
attractors of some dynamical systems \cite{15} are, 
among many others, examples of
occurence of continuous but highly irregular (nondifferentiable) curves
and surfaces.
Velocity
field of a turbulent fluid \cite{26} at low viscosity is a well-known example of
a multifractal function. Ordinary calculus
is inadequate to characterize and handle such curves and surfaces.
Some recent papers~\cite{3,4,5,6} indicate a connection 
between fractional calculus~\cite{7,8,9}
and fractal structure \cite{2,10} or fractal processes \cite{11,22,23}.
However the precise nature of the 
connection between the dimension of the graph of a fractal curve 
and  fractional differentiability
properties was recognized only recently.
A new notion of {\it local fractional derivative} (LFD) was
introduced~\cite{41,81} to study fractional differentiability properties of 
irregular functions. 
An intresting observation of this work was that the local H\"older exponent/
dimension was related to the maximum order for which LFD existed.
In this paper we briefly review the concept of LFD
as applied to a function of one variable
and generalize it for  functions of several variables.

An irregular function of one variable is
 best characterized locally by a {\it H\"older exponent}. We will use the following general definition of the H\"older
exponent which has been used by
various authors~\cite{27,24} recently.
The exponent $h(y)$ of a function $f$
at $y$ is given by $h$ 
such that there exists a polynomial
$P_n(x)$ of order $n$, where $n$ is the largest integer smaller than
$h$, and
\begin{eqnarray}
\vert f(x) - P_n(x-y) \vert = O(\vert x-y \vert^{h}),
\end{eqnarray}
for $x$ in the neighbourhood of $y$.
This definition serves to classify the behavior of the function at $y$.

Fractional calculus~\cite{7,8,9} is a study which deals  with generalization 
of differentiation and integration to fractional orders.
There are number of ways (not neccesarily equivalent) of
 defining fractional derivatives and integrations.
We use the Riemann-Liouville definition.
We begin by recalling the Riemann-Liouville definition of fractional integral
of a real function, which is
given by~\cite{7,9}
\begin{eqnarray}
{{d^qf(x)}\over{[d(x-a)]^q}}={1\over\Gamma(-q)}{\int_a^x{{f(y)}\over{(x-y)^{q+1}}}}dy
\;\;\;{\rm for}\;\;\;q<0,\label{def1}
\end{eqnarray}
where the lower limit $a$ is some real number and of the fractional derivative
\begin{eqnarray}
{{d^qf(x)}\over{[d(x-a)]^q}}={1\over\Gamma(n-q)}{d^n\over{dx^n}}{\int_a^x{{f(y)}\over{(x-y)^{q-n+1}}}}dy
\label{def2}
\end{eqnarray}
for $n-1<q<n$.
Fractional derivative of a simple function $f(x)=x^p\;\;\;p>-1$ 
is given by~\cite{7,9}
\begin{eqnarray}
{{d^qx^p}\over {d x^q}} = {\Gamma(p+1) \over {\Gamma(p-q+1)}} x^{p-q}\;\;\;
{\rm for}\;\;\;p>-1.\label{xp}
\end{eqnarray}
Further the fractional derivative has the 
interesting property (see ref~\cite{7}), viz,
\begin{eqnarray}
{d^qf(\beta x)\over{d x^q}}={{\beta}^q{d^qf(\beta x)\over{d(\beta x)^q}}}
\end{eqnarray}
which makes it suitable for the study of scaling.
Note the nonlocal character of the fractional integral and derivative 
in the equations (\ref{def1}) and (\ref{def2}) respectively. Also it is clear
from equation~(\ref{xp}) that unlike in the case of integer derivatives
the fractional derivative of a constant is not zero in general.
These two features make the extraction of scaling information somewhat
difficult. The problems were overcome by the introduction of LFD in \cite{41}.
In the following section we briefly review the notion of LFD 
for the real valued functions of real variable.
In the section \ref{higherd} we generalize this definition
to real valued functions of many variables and demonstrate it with
the help of some simple examples. 

\section{Local Fractional Differentiability} \label{secdefn}

Unfortunately, as noted in section \ref{secintro}, fractional
derivatives are not local in nature.
On the other hand it is desirable and occasionally crucial to have
local character in wide range of applications ranging from the structure
of differentiable manifolds to various physical models. 
Secondly the fractional derivative of a constant is not zero, consequently 
 the magnitude of the fractional derivative changes with
the addition of a constant.
The appropriate new notion of fractional differentiability must bypass the
hindrance due to these two properties.
 These difficulties were remedied by introducing the notion
LFD in~\cite{41} as follows:
\begin{defn} \label{defLFD1}
If, for a function $f:[0,1]\rightarrow I\!\!R$, the  limit 
\begin{eqnarray}
I\!\!D^qf(y) = 
{\lim_{x\rightarrow y} {{d^q(f(x)-f(y))}\over{d(x-y)^q}}}\label{defloc}
\end{eqnarray}
exists and is finite, then we say that the {\it local fractional derivative} (LFD) 
of order $q$ $(0<q<1)$, at $x=y$, 
exists. 
\end{defn}
In the above definition the lower limit $y$ is treated as a constant.
The subtraction of $f(y)$ corrects for the fact that the fractional
derivative of a constant is not zero. Whereas the limit as $x\rightarrow y$
is taken to remove the nonlocal content.
Advantage of defining local fractional derivative in this manner lies
in its local nature and hence allowing the study of pointwise behaviour
of functions. This will be seen more clearly 
after the development of Taylor series below. 
\begin{defn} \label{defco}
We define {\it critical order} $\alpha$, at $y$, as
$$
\alpha(y) = Sup \{q \;\vert\; {\rm {all\;LFDs\; of\; order\; less\; than\;}} q{{\rm\; exist\; at}\;y}\}. 
$$
\end{defn}
These definitions were subsequently generalized~\cite{81} for $q > 1$ as
follows. 
\begin{defn} \label{defLFD2}
If, for a function $f:[0,1]\rightarrow I\!\!R$, the  limit 
\begin{eqnarray}
I\!\!D^qf(y) =  {\lim_{x\rightarrow y}}
{{d^q(f(x)-\sum_{n=0}^N{f^{(n)}(y)\over\Gamma(n+1)}(x-y)^n)}
\over{[d(x-y)]^q}} \label{deflocg}
\end{eqnarray}
exists and is finite,
where $N$ is the largest integer for which $N^{th}$ derivative of $f(x)$ at
$y$ exists and is finite, then we say that the {\it local fractional
derivative} (LFD) of order $q$ $(N<q\leq N+1)$, at $x=y$, 
exists. 
\end{defn}
We consider this as the generalization of the local derivative for order 
greater than one. 
Note that when $q$ is a positive integer ordinary derivatives are recovered.
The definition of the critical order remains the same
since, for $q<1$, (\ref{defloc}) and (\ref{deflocg}) agree.
This definition extends the applicability of LFD to $C^1-$functions 
which are still irregular due to the nonexistence of some 
higher order derivative (i.e. belong to class $C^\gamma$, $\gamma > 1$).
For example  the critical order of 
$f(x)=a+bx+c\vert x\vert ^{\gamma}$, $\gamma > 1$, at origin,
according to definitions \ref{defco} and \ref{defLFD2} is $\gamma$.
It is clear that the critical order of the $C^\infty$ function is $\infty$.

In~\cite{41} it was shown that the Weierstrass nowhere differentiable
function, given by
\begin{eqnarray}
W_{\lambda}(t) = \sum_{k=1}^{\infty} {\lambda}^{(s-2)k} 
sin{\lambda}^kt,\label{Weier}
\end{eqnarray}
where $\lambda>1$, $1<s<2$ and $t$ real,
is continuously locally fractional differentiable for orders below
$2-s$ and not for orders between $2-s$ and one. This implies that
the critical order of this function is $2-s$ at all points.
Interesting consequence of this result is the relation between
box dimension $s$ \cite{10} of the 
graph of the function and the critical order. This
result has for the first time given a direct
relation between the differentiability
property and the dimension. In fact this 
observation was consolidated into a general
result showing equivalence between critical order and the local H\"older
exponent of any continuous function. The LFD was further shown to be useful
in the study of pointwise behaviour of multifractal functions and 
for unmasking the singularities masked by stronger singularities.

Whenever it is required to distinguish between limits from right and left sides
we can write the definition for LFD in the following form.
\begin{eqnarray}
I\!\!D_{\pm}^qf(y) =  {\lim_{x\rightarrow y^{\pm}}}
{{d^q\widetilde{F}_N(x,y)}
\over{[d\pm(x-y)]^q}} \label{deflocg+}
\end{eqnarray}

The importance of the notion of LFD lies in the fact that it appears
naturally in the
fractional Taylor's series with a remainder term for a real function $f$,
given by, 
\begin{eqnarray}
f(x) = \sum_{n=0}^{N}{f^{(n)}(y)\over{\Gamma(n+1)}}\Delta^n
 + {I\!\!D^q_+f(y)\over \Gamma(q+1)} \Delta^q + R_q(y,\Delta) \label{taylor}
\end{eqnarray}
where $x-y=\Delta > 0$ and $R_q(y,\Delta)$ is a remainder given by
\begin{eqnarray}
R_q(y,\Delta) = {1\over\Gamma(q+1)}\int_0^{\Delta} 
{dF(y,t;q,N)\over{dt}}{(\Delta-t)^q}dt
\end{eqnarray}
and
\begin{eqnarray}
F(y,\Delta;q,N) = 
{d^q(f(x)-\sum_{n=0}^{N}{f^{(n)}(y)\over{\Gamma(n+1)}}\Delta^n)
\over{[d\Delta]^q}} \label{F}
\end{eqnarray}
We note that the local fractional derivative as defined above
(not just fractional derivative), along with the first $N$ derivatives,
 provides an approximation
of $f(x)$  
in the vicinity of $y$. 
 We further remark that the terms
on the RHS of eqn(\ref{taylor}) are nontrivial and finite only in the case 
when $q$ equals $\alpha$, the critical order.
Osler~\cite{21} constructed a fractional Taylor
series using usual (Riemann-Liouville)
  fractional derivatives which was applicable only
to analytic function.
 Further Osler's
formulation involves terms with negative orders also and hence is not suitable
for approximating schemes. When $\Delta < 0$, a similar expansion can be written
for $I\!\!D^q_-f(y)$ by replacing $\Delta$ by $-\Delta$.

When $0<q<1$ we get as a special case
\begin{eqnarray}
f(x) = f(y)
 + {I\!\!D^qf(y)\over \Gamma(q+1)} (x-y)^q + Remainder \label{taylor01}
\end{eqnarray}
provided the RHS exists.
If we set $q$  equal to
one in equation (\ref{taylor01}) one gets
the equation of the tangent. 
All the curves passing through a point $y$ and having same the tangent,
form an equivalence class (which is modelled by a linear behavior). 
Analogously all the functions (curves) with the same critical order $\alpha$
and the same $I\!\!D^{\alpha}$
will form an equivalence class modeled by the power law $x^{\alpha}$.  
This is how one may
generalize the geometric interpretation of derivatives in terms of tangents.  
This observation is useful in the  approximation of an irregular 
function by a piecewise smooth (scaling) function.
One may recognize the introduction of such equivalence classes
as a starting point for
fractional differential geometry.

\section{Generalization to higher dimensional functions} \label{higherd}
The definition of the Local fractional derivative can be generalized
for higher dimensional functions in the following manner.

Consider a function $f: I\!\!R^n \rightarrow I\!\!R$. We define
\begin{eqnarray}
\Phi({\bf y},t) = f({\bf y}+{\bf v}t) - f({\bf y}),\;\;\;
{\bf v} \in I\!\!R^n,\;\;\;t\in I\!\!R.
\end{eqnarray}
Then the directional-LFD of $f$ at ${\bf y}$ 
of order $q$, $0<q<1$, in the direction ${\bf v}$ is given 
(provided it exists) by
\begin{eqnarray}
I\!\!D^q_{\bf v}f({\bf y}) = {d^q{\Phi}({\bf y},t) \over{dt^q}}\vert_{t=0} 
\label{defloch}
\end{eqnarray}
where the RHS involves the usual fractional 
derivative of equation (\ref{def2}). The directional-LFDs along the unit vector 
${\bf e}_i$ will be called $i^{\rm th}$ partial-LFD.

Now let us consider two examples.

\noindent
{\bf Example 1:} Let
\begin{eqnarray}
W_{\lambda}({\bf x}) = W_{\lambda}(x,y) = \sum_{k=1}^{\infty} 
{\lambda}^{(s-2)k} sin{\lambda}^k(x+y),
\end{eqnarray}
with $\lambda>1$ and $1<s<2$.
Let ${\bf v} = (v_x, v_y)$ be a unit 2-vector. Then
\begin{eqnarray}
W_{\lambda}({\bf x+{\bf v}t}) &=& W_{\lambda}(x+v_xt,y+v_yt) \nonumber \\
&=& \sum_{k=1}^{\infty} 
{\lambda}^{(s-2)k} sin{\lambda}^k(x+v_xt+y+v_yt).\nonumber
\end{eqnarray}
\begin{eqnarray}
\Phi({\bf x},t) &=& W_{\lambda}({\bf x+{\bf v}t}) - W_{\lambda}({\bf x})
\nonumber\\
&=& \sum_{k=1}^{\infty} 
{\lambda}^{(s-2)k} [sin{\lambda}^k(x+v_xt+y+v_yt) \nonumber \\
&&\;\;\;\;\;\;\;\;\;\;\;\;\;\;\;\;\;\;\; - sin{\lambda}^k(x+y)]
\nonumber
\end{eqnarray}
If we choose $y=0$, i.e., we examine fractional differentiability at a point on
x axis and if we choose $v_y=0$, i.e., we are looking at LFD in the direction
of increasing $x$ then we get
\begin{eqnarray}
\Phi({\bf x},t) &=& \sum_{k=1}^{\infty} 
{\lambda}^{(s-2)k} [sin{\lambda}^k(x+v_xt) - sin{\lambda}^k(x)] \nonumber
\end{eqnarray}
which is known \cite{41} to have critical order $2-s$.

If we keep $y=0$ but keep $v_x$ and $v_y$ non-zero, i.e., if we examine
fractional differentiability at a point on $x$-axis but in any direction
then we get
\begin{eqnarray}
\Phi({\bf x},t) &=& \sum_{k=1}^{\infty} 
{\lambda}^{(s-2)k} [sin{\lambda}^k(x+v_xt+v_yt) - sin{\lambda}^k(x)] \nonumber
\end{eqnarray}
This again has critical order $2-s$, except when $v_x=-v_y$
(in which case it is $\infty$ since the function $\Phi$ identically
vanishes).

In fact it can be shown that the given function has $2-s$ as critical order
at any point and in any direction except the direction with $v_x=-v_y$.

\noindent
{\bf Example 2:}
Another example we consider is
\begin{eqnarray}
W_{\lambda}({\bf x}) = W_{\lambda}(x,y) = \sum_{k=1}^{\infty} 
{\lambda}^{(s-2)k} sin{\lambda}^k(xy),
\end{eqnarray}
where $\lambda>1$ and $1<s<2$.
\begin{eqnarray}
W_{\lambda}({\bf x+{\bf v}t}) &=& W_{\lambda}(x+v_xt,y+v_yt)  \nonumber \\
&=& \sum_{k=1}^{\infty} 
{\lambda}^{(s-2)k} sin({\lambda}^k(x+v_xt)(y+v_yt)).\nonumber
\end{eqnarray}
\begin{eqnarray}
\Phi({\bf x},t) &=& W_{\lambda}({\bf x+{\bf v}t}) - W_{\lambda}({\bf x})
\nonumber\\
&=& \sum_{k=1}^{\infty} 
{\lambda}^{(s-2)k} [sin({\lambda}^k(xy+yv_xt+xv_yt+v_xv_yt^2))\nonumber \\
&&\;\;\;\;\;\;\;\;\;\;\;\;\;\;\;\;\;\;\;\; - 
sin({\lambda}^k(xy))]\nonumber
\end{eqnarray}
Now if we choose $y=0$ and $v_y=0$ then the critical order is $\infty$.

If we choose $y=0$ but $v_x$ and $v_y$ non-zero then we get
\begin{eqnarray}
\Phi({\bf x},t)&=& \sum_{k=1}^{\infty} 
{\lambda}^{(s-2)k} [sin({\lambda}^k(xv_yt+v_xv_yt^2))] \nonumber
\end{eqnarray}
Therefore using results of~\cite{41}
the critical order along any other direction ${\bf v}$,
at a point on $x$-axis is seen to be $2-s$.

\section{Conclusions}
The usefulness of the notion of LFD was pointed out in \cite{41,81}
where the considerations were restricted to functions of one variable only.
It allows us to quantify the loss of differentiability
of fractal and multifractal functions. 
The larger the irregularity of the functions the smaller
is the extent of differentiability 
and smaller is the value of the H\"older exponent.
Local Taylor series expansions provide a way to approximate irregular functions
by a piecewise scaling functions. In the present
paper we have demonstrated that it is
possible to carry the same theme even in the multidimensional case.
In particular, the H\"older exponents in any direction are related to
the critical order of the corresponding directional-LFD.
We note that, whereas a one dimensional result is useful in studying
irregular signals, the results here may have utility in image processing
where one can characterize and classify singularities in the image data.
We note that it is possible to write a multivariable fractional Taylor
series which can be used for approximations and modelling of 
multivariable multiscaling functions. This development
will be reported elsewhere.
Further these considerations provide a way for formulating fractional
differential geometry for fractal surfaces.

One of the author (KMK) would like to thank 
CSIR (India) and the other author (ADG) would 
like to thank DBT (India) for financial assistance.

\end{multicols}
\end{document}